\documentclass[ twocolumn]{aastex631}

\newcommand{\kms}{$\rm km\, s^{-1}$\,} 
\newcommand{\lya}{Ly$\alpha$\,}

\newcommand{\HeII}{\mbox{He\,{\sc ii}}}
\newcommand{\HI}{\mbox{H\,{\sc i}}}

\newcommand{\CIV}{\mbox{C\,{\sc iv}}}

\newcommand{\CII}{\mbox{C\,{\sc ii}}}

\newcommand{\CIII}{\mbox{C\,{\sc iii}}}

\newcommand{\SiII}{\mbox{Si\,{\sc ii}}}
\newcommand{\SiIII}{\mbox{Si\,{\sc iii}}}
\newcommand{\SiIV}{\mbox{Si\,{\sc iv}}}

\newcommand{\MgII}{\mbox{Mg\,{\sc ii}}}

\newcommand{\AlIII}{\mbox{Al\,{\sc iii}}}

\newcommand{\logN}{$\log_{10} N/{\rm cm^{-2}}$} 

\newcommand{\ang}{\mbox{\normalfont\AA}}
\newcommand{\NHI}{$N_{\rm H \textsc{i}}$}

\newcommand{\nh}{$n_{\rm H}$}
\newcommand{\met}{$[X/H]$}

\newcommand{\Rev}[1]{\textcolor{black}{#1}}

\begin{document}

\title{MUSEQuBES: Unveiling Cosmic Web Filaments at $z\approx3.6$ through Dual Absorption and Emission Line Analysis}

\correspondingauthor{Eshita Banerjee, Sowgat Muzahid}
\email{eshitaban18@iucaa.in, sowgat@iucaa.in} 

\author[0009-0002-7382-3078]{Eshita Banerjee}
\affiliation{IUCAA, Post Bag 04, Ganeshkhind, Pune, India, 411007} 

\author[0000-0003-3938-8762]{Sowgat Muzahid}
\affiliation{IUCAA, Post Bag 04, Ganeshkhind, Pune, India, 411007} 

\author{Joop Schaye}
\affiliation{Leiden Observatory, Leiden University, P.O. Box 9513, NL-2300 AA Leiden, the Netherlands}

\author{Sebastiano Cantalupo}
\affiliation{Department of Physics, University of Milan Bicocca, Piazza della Scienza 3, I-20126 Milano, Italy}

\author{Sean D. Johnson}
\affiliation{Department of Astronomy, University of Michigan, 1085 S. University, Ann Arbor, MI 48109, USA}



\begin{abstract}

According to modern cosmological models, galaxies are embedded within cosmic filaments, which supply a continuous flow of pristine gas, fueling star formation and driving their evolution. However, due to their low density, the direct detection of diffuse gas in cosmic filaments remains elusive. 
Here, we report the discovery of an extremely metal-poor ($[ X/H] \approx -3.7$), low-density ($\log_{10} n_{\rm H}/{\rm cm^{-3}} \approx -4$, corresponding to an overdensity of $\approx 5$) partial Lyman limit system (pLLS) at $z\approx3.577$ along the quasar sightline Q1317--0507, probing cosmic filaments. Additionally, two other low-metallicity (\met$\lesssim -2$) absorption systems are detected at similar redshifts, one of which is also a pLLS. 
\Rev{VLT/MUSE observations reveal a significant overdensity of \lya\ emitters (LAEs) associated with these absorbers.}  
The spatial distribution of the LAEs strongly suggests the presence of an underlying filamentary structure. 
This is further supported by the detection of a large \lya\ emitting nebula with a surface brightness of $\geq 10^{-19}~\rm erg~cm^{-2}~s^{-1}~arcsec^{-2}$, with a maximum projected linear size of $\approx 260$~pkpc extending along the LAEs.
\Rev{This is the first detection of giant \lya\ emission tracing cosmic filaments, linked to normal galaxies and  likely powered by in-situ recombination.}

\end{abstract}



\section{Introduction} \label{sec:intro}

In the current cosmological framework, galaxies emerge within the dense intersections of the cosmic web—a large-scale network of dark matter halos and filaments that span the universe. These structures channel gas from the intergalactic medium (IGM) into dark matter halos, where it eventually cools, triggering star formation. However, detecting these emission from the gas in elusive filaments is challenging due to their low densities. 

Recent advancements in \Rev{integral field units (IFUs)} with large fields of view, like MUSE \citep{bacon2010muse}, have revolutionized our ability to detect these filament-like structures, glowing in \lya-emission at high redshifts \citep{Fumagalli2016, Bacon2021, Johnson_2022, Tornotti2024}. These observations offer new insights into gas flow from the IGM into galaxies, particularly through ``cold-mode accretion'' \citep[e.g.,][]{Keres_2005}, where gas is funneled into galaxies via narrow, dense filaments. This process significantly contributes to the optically thick gas associated with Lyman-limit systems (LLSs: $\log_{10}$(\NHI)~$>17.2$) \citep[see,][]{Fumagalli2011, van_de_voort_2012}.

\cite{Fumagalli2013} have shown that while gas in galaxy halos can account for all LLSs at $z < 3$, at $z \gtrsim 3.5$, the contribution of the IGM to LLSs becomes pronounced, as the overdensities associated with these systems decrease \citep[see,][]{Schaye_2001} and the extragalactic UV background (UVB) weakens, enhancing gas shielding. Consequently, LLSs are considered effective tracers of cold-stream inflows onto galaxies, often identified by their low metallicity \citep[e.g.,][]{Ribaudo2011, Crighton2013} or filamentary morphology \citep[e.g.,][]{Fumagalli2016}. At $z \approx 3$, only a small fraction ($\approx 18\%$) of LLSs and partial-LLSs (pLLSs: $16.2< \log_{10}$(\NHI)~$<17.2$) are extremely metal-poor, with metallicity being \met$<-3$ \citep{Lehner2016, kodiaq_z, Lofthouse_2023}.

Interestingly, in our MUSEQuBES survey, \Rev{we identified an overdensity of \lya\ emitters (LAEs)} at $z \approx 3.577$, consisting of seven LAEs arranged in an almost linear configuration. Suspecting a filament connecting these LAEs, we explored potential inflow signatures by modeling absorbers probed by a background quasar and searched for extended emission around this structure. This investigation revealed a low-metallicity absorption system and a coincident giant \lya\ nebula. This letter is organized as follows: section~\ref{sec:data} introduces our data; section~\ref{sec:result} presents absorption measurements and modeling, and finally, we summarize our study and discuss the results in section~\ref{sec:discussion}. We adopt a flat $\Lambda$CDM cosmology with $H_0$ = $70\,\rm km\, s^{-1}\, Mpc^{-1}$, $\Omega_{\rm M} = 0.3$ and $\Omega_\Lambda = 0.7$. Metallicity is expressed as $\log_{10} (Z/ \rm Z_{\odot})\equiv$\met, where $\rm Z_{\odot}$ is the solar metallicity ($=0.013$;  see \cite{Grevesse2012}). Distances are in physical kpc (hereafter, pkpc) unless specified otherwise.


\section{Data}
\label{sec:data}

The LAE overdensity analyzed in this study is detected toward the quasar Q1317$-$0507, observed as part of the MUSEQuBES survey \citep{Muzahid_2020, Muzahid_2021,Banerjee2023, Banerjee2024}. We obtained 10 hours of on-source \Rev{VLT/MUSE} observations with an effective seeing of $< 0.6^{''}$. The final data cube has a spatial sampling of $0.2^{''} \times 0.2^{''}$ per pixel, and a spectral resolution of $\approx 3600$ (FWHM $\approx86$~\kms) in the optical range (4750–9350 \ang). The data reduction process is comprehensively described in \cite{Muzahid_2021}.

Complementary to the MUSE data, we utilized a high-resolution optical spectrum of the quasar from VLT/UVES ($R \approx 45,000$), sourced from the SQUAD database \citep{Murphy2019}. The coadded and continuum-normalized spectrum provides a median signal-to-noise ratio (SNR) of $35$ within the \lya-forest region and $80$ redward of the quasar's \lya\ emission. Additionally, we incorporated near-infrared data from \Rev{VLT/X-shooter}, covering 1000-2480~nm with a spectral resolution of $R \approx 5300$ and a median SNR of $\approx 35$. This spectrum, along with its best-fitting continuum, were retrieved from the ESO data archive \citep{Lopez2016}.


\section{Analysis and Results}
\label{sec:result}

\begin{figure}
    \centering
    \includegraphics[width=0.5\textwidth]{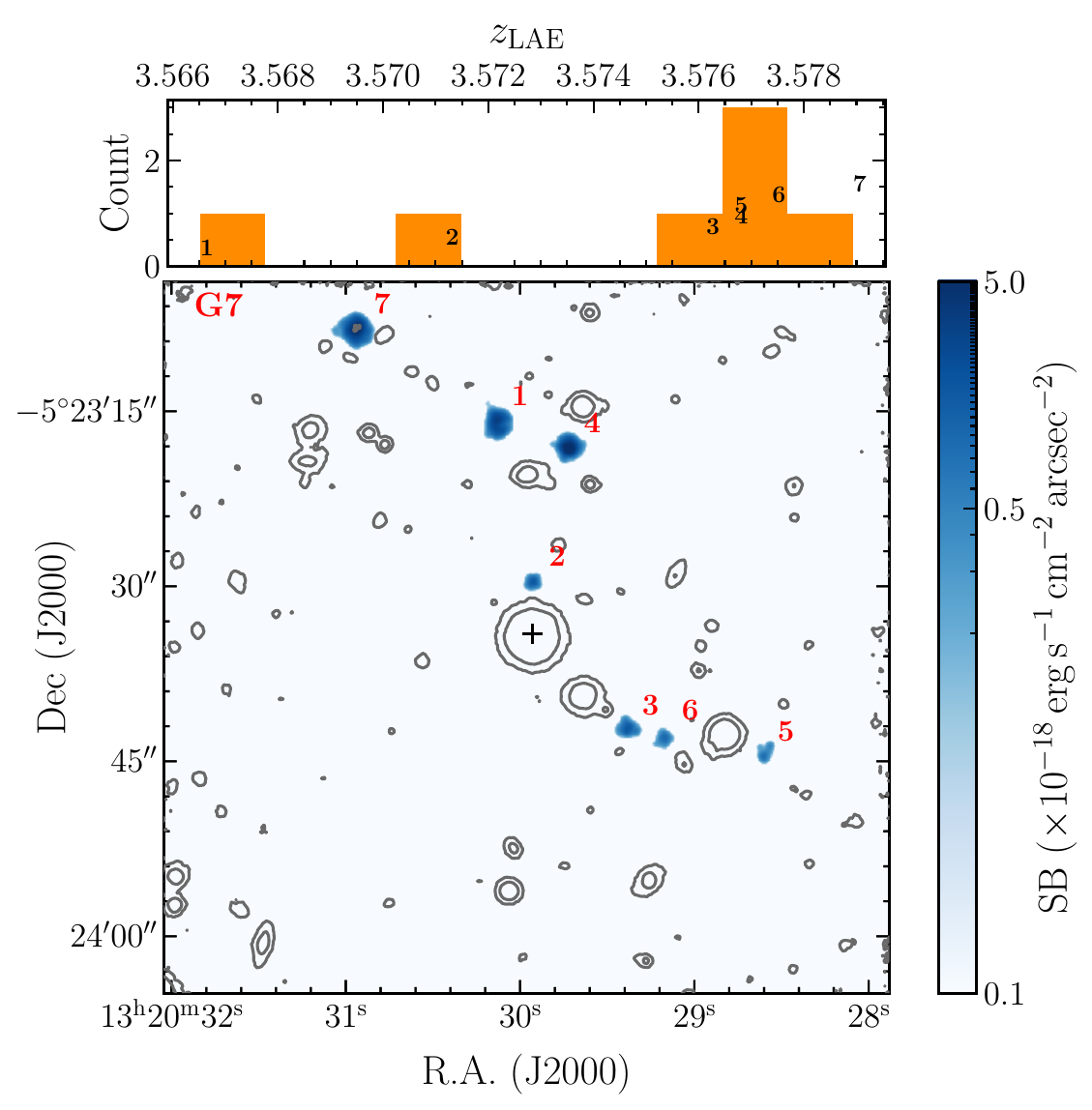} 
    \caption{The optimally extracted \lya\ surface brightness maps of the 7 LAEs (G7) within the MUSE FOV centered on the quasar Q1317$-$0507 (marked by the ``$+$'' sign). The pixels within the 3D segmentation map for each LAE are combined and projected onto the image, with the gray contours representing the 5 and 25~$\sigma$ from the mean flux levels of the continuum-bright objects. A Gaussian smoothing function with $\sigma=0.2''$ ($\equiv1$ pixel) has been applied to enhance visual clarity of the SB map. The histogram in top panel displays the redshift distribution of the LAEs. The object IDs are indicated beside each LAEs as well as in the histogram plot.}
    \label{fig:group_img}
\end{figure}


\cite{Muzahid_2020} identified 22 LAEs in the MUSE field centered on the background quasar Q1317$-$0507 ($z_{\rm qso}=3.7$) in the redshift range $2.9<z<3.6$. These LAEs were detected based on their \lya\ emission lines, which typically show offsets of hundreds of \kms\ from the systemic redshifts \citep[e.g.,][]{Steidel_2010, Rakic+11, Shibuya_2014, Verhamme18}. The \lya\ redshifts were corrected using the empirical relation from \cite{Muzahid_2020}. A friends-of-friends algorithm, using a linking velocity\footnote{\Rev{earlier, \cite{Muzahid_2021} also used the similar velocity window for defining galaxy-groups.}} of $500$~\kms\ along the line of sight (LOS), identified a galaxy overdensity with 7 LAEs at $z \approx 3.57$, making it the most LAE-rich system in the MUSEQuBES sample.

Figure~\ref{fig:group_img} shows the optimally extracted \lya\ surface brightness (SB) map of this overdense region (hereafter, G7). The redshifts of the seven LAEs range from $z \approx 3.566$ to $3.578$. The LAE closest to the quasar-sightline is Id:2, at a transverse distance of $34$~pkpc, followed by Id:3 at $91$~pkpc. The other LAEs are located beyond $100$~pkpc, with the farthest at $220$~pkpc. The redshift histogram reveals that five of the seven LAEs (excluding Id:1 and Id:2) are tightly clustered at $z \approx 3.577$, which is $\approx 8000$~\kms\ or 20~pMpc from the background quasar.


\begin{figure*}
    \centering
    \includegraphics[width=\textwidth]{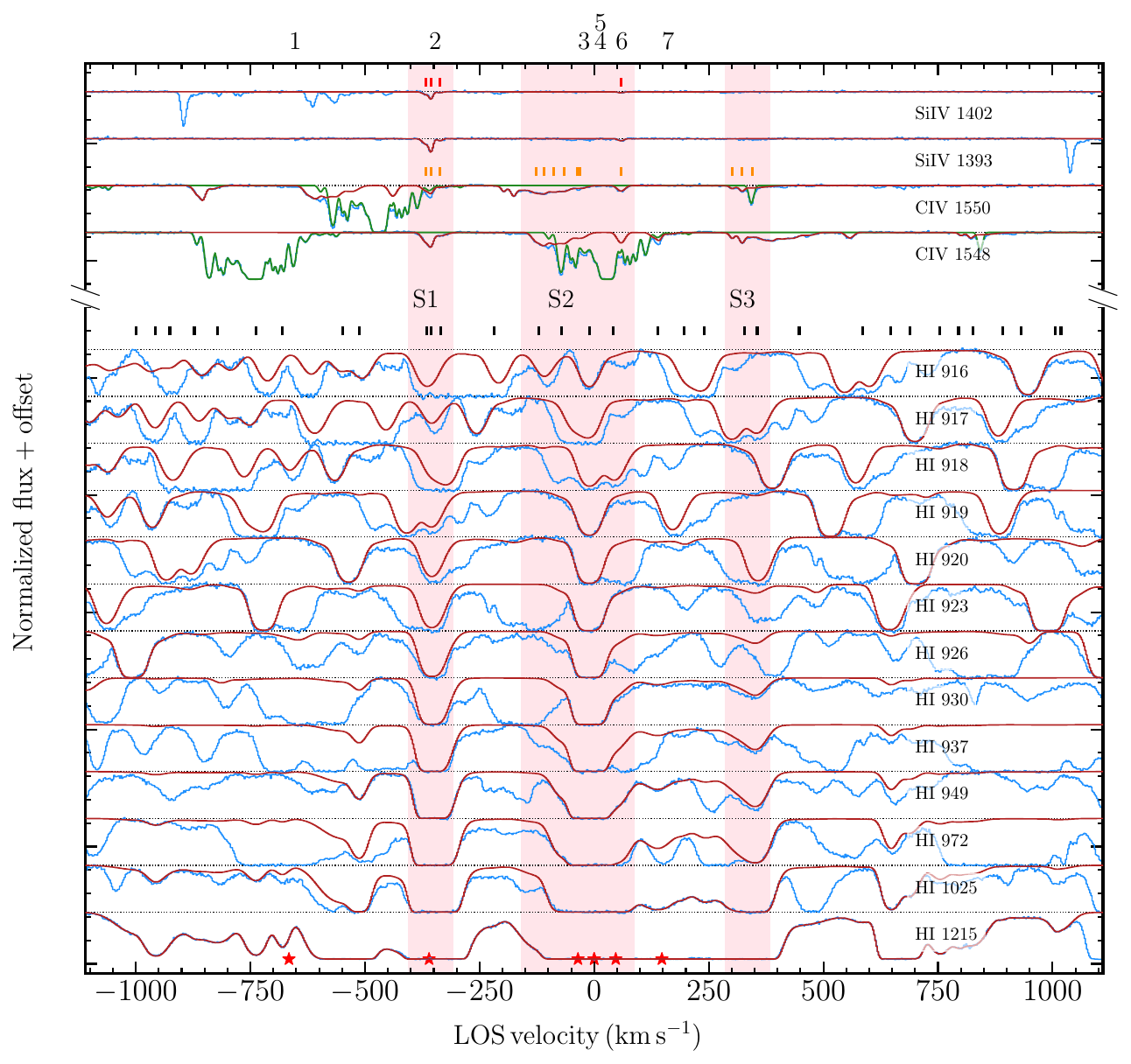}
    \caption{Velocity plot of the G7 system, showing several Lyman-series lines (arranged in descending order of wavelength from bottom to top) along with metal lines detected in the system. The observed spectrum is shown in blue, with the best-fitting model curve from {\sc vpfit} in red. Fits for contaminating transitions are displayed in green. The black, orange and red ticks mark the positions of individual \HI, \CIV\ and \SiIV\ components, respectively. For visual clarity, the normalized fluxes of each line are shifted vertically. The positions of the seven LAEs are marked by red stars at the bottom and their respective IDs are written on top, with $\Delta v=0$ corresponding to their median redshift of $z=3.577$. The total column densities measured within the three pink shaded regions (S1, S2 and S3) are used to construct photoionization models.
    }
    \label{fig:veloplot}
\end{figure*}

\subsection{Measurements of absorption lines associated with G7}

Fig.~\ref{fig:veloplot} shows the velocity plot for the Lyman-series lines and metal transitions associated with the G7 system. The seven LAEs are marked by red stars, with $\Delta v = 0$ corresponding to their median redshift of $z = 3.577$. To constrain the \HI\ absorber parameters, we simultaneously fitted the Lyman-series lines, from \lya\ to \HI-$\lambda$916, using the Voigt profile fitting software {\sc vpfit} \citep{Carswell2014}. This software minimizes $\chi^2$ to determine the best-fitting redshift ($z$), Doppler parameter ($b$), and column density ($N$) of the absorbers. Strong, un-fitted absorption in higher-order lines is contamination, as evident from the lack of stronger absorption in \lya\ at similar velocities. This underscores the need for simultaneous fitting of all Lyman-series lines. We identified over 30 \HI\ components within $\pm1000$~\kms, including two pLLSs at $-60$ and $-300$~\kms\ with \HI\ column densities of \logN= 16.7 and 16.3, respectively.

Next, we searched for metal transitions associated with G7 within the same velocity range. We detected metal absorption corresponding to \HI\ absorbers at approximately $-300$, $-60$, and $300$~\kms, which we labeled as S1, S2, and S3, respectively. The highlighted velocity ranges used to associate aligned transitions were based on the structure of the detected metal absorption lines. \CIV\ absorption was observed in all three systems, while \SiIV\ was detected in S1 and S2. No other metal transitions were detected within this range. For non-detections, we calculated $3\sigma$ limiting column densities using the $3\sigma$ limiting equivalent width \citep{Hellsten1998}, assuming the linear part of the curve of growth.

When fitting the aligned \CIV\ and \SiIV\ transitions, we tied their redshifts. However, the \CIV1548 line for S2 and \CIV1550 for S1 are contaminated by \MgII\ absorption from $z = 1.52$, while \CIV1550 in S3 is affected by a $z = 2.83$ \AlIII\ line. To accurately measure the metal absorption parameters, we fitted these contaminating lines as well. The $N$ and $b$ of these blended components are reliably constrained because the corresponding unblended, unsaturated doublet lines provide accurate measurements. We also excluded transitions like \CIII\ and \SiIII\ due to heavy contamination from the \lya\ forest.

Among the metal transitions, we identified four pairs of components (three from S1 and one from S2) where \CIV\ and \SiIV\ are aligned in redshift. By analyzing the $b$-parameters of these components, we separated the contributions from temperature ($T$) and turbulent velocity ($v_{\rm turb}$) in the medium using the relation: $b^2 = v_{\rm turb}^2 + \frac{2k_{\rm B}T}{m_{\rm ion}}$. Here, $m_{\rm ion}$ is the mass of ion, $k_{\rm B}$ is the Boltzmann constant and $v_{\rm turb}$ is the turbulent velocity contributing to the non-thermal broadening. The resulting median temperature of the medium is $\rm{log_{10}}\,(T/\rm{K}) \approx 4.7\pm0.2$. While this method can also be applied using a metal ion and its associated \HI\ absorber, it was not feasible here due to the presence of multiple metal absorption components associated with a single \HI\ absorber.

\subsection{Photoionization model}

\begin{table}
    \centering
    \caption{Range of the parameters used for \textsc{Cloudy}}
    \begin{tabular}{cccc}
    \hline
    Parameter & Minimum & Maximum  & Interval \\
    \hline
    $\log_{10} N_{\rm HI}/\rm cm^{-2}$ & 12.5 & 20.5 & 0.25 \\
    $z$  & 2.75 & 4 & 0.25 \\
    \met & -4.0 & 1.0 & 0.25 \\
    $\log_{10} n_{\rm H}/\rm cm^{-3}$  & -4.5 & 0.0 & 0.25 \\
    \hline
    \end{tabular}
    \label{tab:cloudy_params}
    Note:-- For the Bayesian inference code, we have used interpolation to obtain intermediate values.
\end{table}

High column density \HI\ absorbers, such as pLLS and LLS, are typically photoionized across redshifts \citep{Crighton2015, Fumagalli2016, Prochaska2017, CCCI, kodiaq_z}. 


We employed \textsc{Cloudy} (v-C17, \citealp{Ferland2013}) to compute ionization corrections, assuming a uniform slab of gas with a constant hydrogen density (\nh) and solar elemental abundances \citep{Asplund2009} is in thermal and ionization equilibrium. The incident radiation is assumed to be the redshift-dependent UV background (UVB) given by the \citet[][hereafter, HM05]{Haardt_Madau_2001} model, along with the cosmic microwave background (CMB). It is important to note that for $z > 3$, the UVB models (e.g., HM05, HM12 \cite{{Haardt-Madau2012}} or KS18 \citep{Khaire2019}) show minimal variation in the relevant energy range. The model iterated until it reached the neutral hydrogen column density (\NHI). We did not include dust or grains, assuming all elements are in the gas phase.


\begin{figure}
    \centering
    \includegraphics[width=\linewidth]{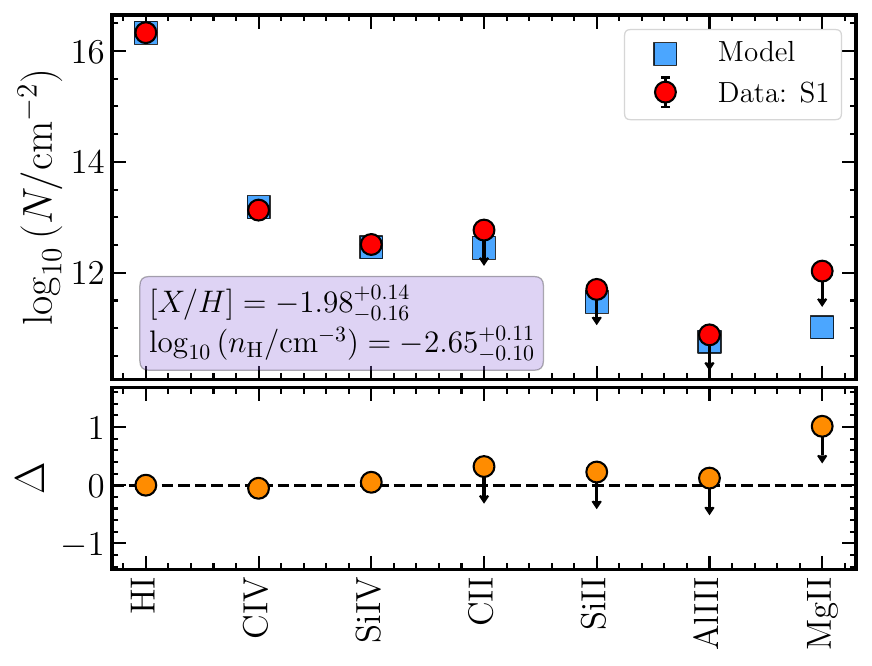} 
    \includegraphics[width=\linewidth]{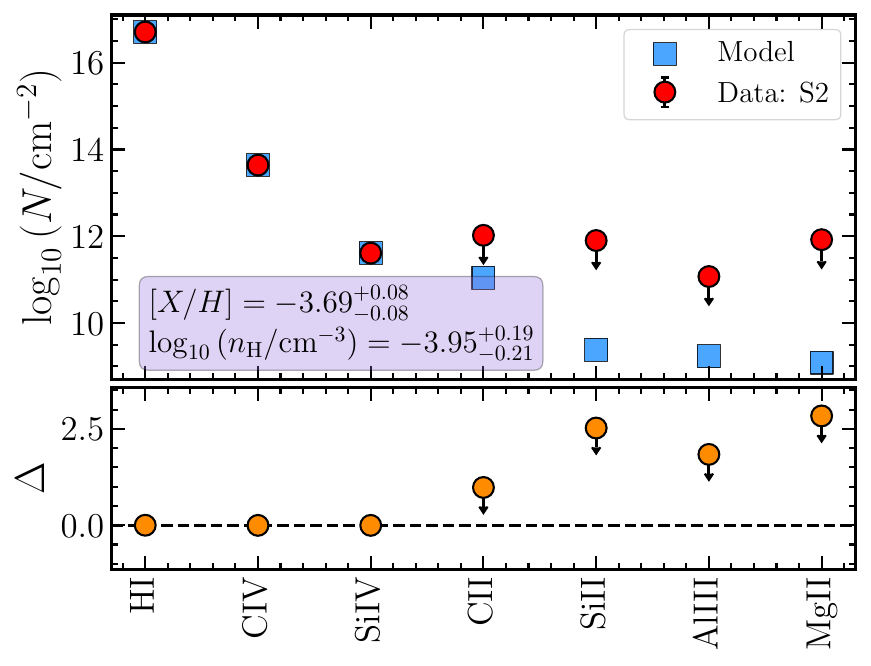}
    \includegraphics[width=\linewidth]{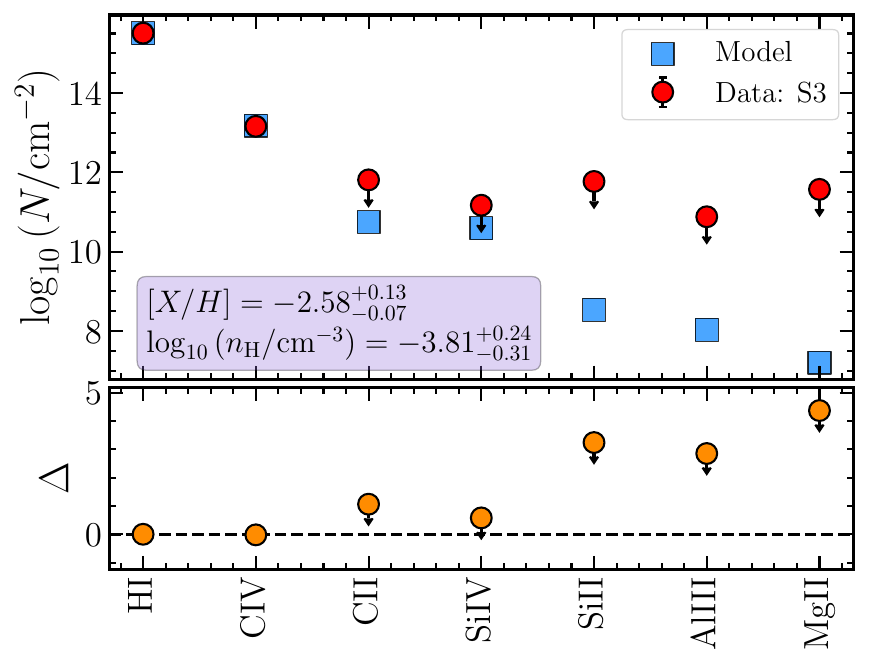}
    \caption{Measured column densities of different transitions associated with S1, S2 and S3 ({\tt top} to {\tt bottom}) are shown in red, with upper limits indicated by downward arrows. Blue squares represent the predicted column densities based on the median values of the model parameters from their respective posterior PDFs. The median values of metallicity and \nh\ along with their $16^{th}$-$84^{th}$ percentile ranges are displayed at the top. The bottom panel shows the residuals ($\Delta \equiv {\rm log_{10}}\,N - {\rm log_{10}}\, N_{\rm model}$).}
    \label{fig:model_result}
\end{figure}

We applied a Bayesian approach to compare the column densities and errors of each ion with a grid of photoionization models to derive the metallicities and densities. 
This method, commonly used in previous studies \citep[e.g.,][]{Fumagalli2011, Crighton2015, CCCI, kodiaq_z}, is effective in producing robust confidence intervals from posterior probability distributions. We employed likelihood functions similar to \cite{Fumagalli2016}. Our model parameters are: (i) neutral hydrogen column density (\NHI), (ii) redshift ($z$), (iii) \Rev{metallicity (\met)}, and (iv) the total hydrogen number density (both neutral and ionized), $n_{\rm H} = n_{\gamma}/U$, where $n_{\gamma}$ is the hydrogen ionizing photon density and $U$ is ionization parameter. The parameter ranges are shown in table~\ref{tab:cloudy_params}. 

We used the nested sampling Monte Carlo algorithm MLFriends implemented in the {\tt UltraNest}\footnote{\url{https://johannesbuchner.github.io/UltraNest/}} package of Python to obtain the posterior probability density functions (PDFs) of the modeling parameters. Gaussian priors were adopted for \NHI\ and $z$ based on \textsc{vpfit} constraints, while flat priors were used for metallicity and \nh\ across the grid's parameter space.


\subsection{Results of Photoionization modeling}
\label{sec:result_val}

Fig.~\ref{fig:model_result} compares the observed column densities with the best-fitting values derived from the medians of the posterior distributions of model parameters for systems S1, S2, and S3. 


System S1 is closely aligned in velocity with LAE Id~2, the galaxy nearest to the quasar sightline. Classified as a pLLS (\logN$=16.3$), it shows \CIV\ and \SiIV\ detections, with upper limits on \CII, \SiII, \AlIII, and \MgII. Bayesian analysis with HM05 (KS18) UVB indicates low metallicity, \met$=-1.98^{+0.14}_{-0.16}\, (-2.22^{+0.17}_{-0.15})$ and density $\log_{10}\,$\nh$/{\rm cm^{-3}}=-2.65^{+0.11}_{-0.10}\, (-2.94^{+0.13}_{-0.11})$.

System S2 lies near the redshift of the clustered LAEs and has the highest \Rev{\logN$=16.7$}, also classified as a pLLS. It shows \CIV\ and weak \SiIV\ detections. The metallicity of S2 is extremely low, with \met$=-3.69^{+0.08}_{-0.08}$ (HM05) or \met$=-4.08^{+0.08}_{-0.08}$ (KS18). The density is $\log_{10}\,$\nh$/{\rm cm^{-3}}=-3.95^{+0.2}_{-0.2}\, (-4.27^{+0.18}_{-0.14})$ for HM05 (KS18).

System S3 has only \CIV\ detected in addition to \HI. Hence, instead of using flat priors, a Gaussian prior on density with $\log_{10}\,$\nh$/{\rm cm^{-3}} \approx -3.5 \pm 0.5$, was applied. This corresponds to the maximum \CIV\ ion-fraction for the given \NHI\ for the metallicities and redshift range included in our grid. This system is also metal-poor, with \met$=-2.58^{+0.13}_{-0.07}\, (-2.94^{+0.16}_{-0.13})$ and $\log_{10}\,$\nh$/{\rm cm^{-3}}=-3.81^{+0.24}_{-0.32}\, (-4.02^{+0.17}_{-0.27})$ using HM05 (KS18) UVB.




\section{Discussion and Conclusion}
\label{sec:discussion}

\subsection{The G7 system as a tracer of filamentary structure} 
 

To assess the overdensity of the G7 system, we used the LAE luminosity function (LF) to estimate the expected number of LAEs in a cosmological volume corresponding to that of the G7 system. The LF of \citet{Drake2017} \citep{Herenz2019} predicts only $0.6$ ($0.8$) LAEs with $\log_{10}~ (L_{\rm Ly\alpha}/\rm erg~s^{-1}) \geq 41.4$, which is the lowest detected luminosity in the G7 system. Detecting seven LAEs thus corresponds to a Poisson probability of $3\times10^{-6}~(2\times10^{-5}$ for a mean of 0.8), confirming that this is a highly overdense region.

The projected spatial distribution of G7 member LAEs is notably non-random, forming a near-linear structure (see Fig.~\ref{fig:group_img}). Using a Monte Carlo toy model, we estimated the chance probability of this alignment. By fitting the pixel coordinates of LAEs with a straight line\footnote{using \href{https://scikit-learn.org/dev/modules/generated/sklearn.linear_model.LinearRegression.html}{LinearRegression} class from Scikit-learn.}, we measured the maximum perpendicular distance ($\epsilon$) from the line. Randomly placing seven points in the $324 \times 324$ spaxel$^2$ MUSE FoV, we repeated this process to compute $\epsilon_i$ for \Rev{1000 realizations}. The probability of $\epsilon_i \leq \epsilon$ was found to be $\approx 0.3\%$, indicating that such alignments are extremely rare.

Five of the seven G7 LAEs (excluding Ids-1 and 2) are clustered in LOS velocity at $z \approx 3.577$, closely matching the velocity ($\Delta v \approx -60$~\kms) of the extremely metal-poor system S2 (\met$ = -3.69$). These LAEs lie at a projected distance of $100-200$~pkpc from the quasar-sightline. S2, with $\log_{10}\,$\nh$/{\rm cm^{-3}} = -4.0$ \citep[corresponding to an  overdensity of $\delta \approx 5$;][]{Schaye_2001}, likely originates from cosmic filaments rather than the CGM \citep[e.g.,][]{Crighton2013, Fumagalli2016_fila, Fumagalli2016, Mackenzie_2019}. We therefore investigated whether there are any traces of extended \lya\ emission around this LAE overdensity.


\begin{figure}
    \centering
    \includegraphics[width=1.1\linewidth]{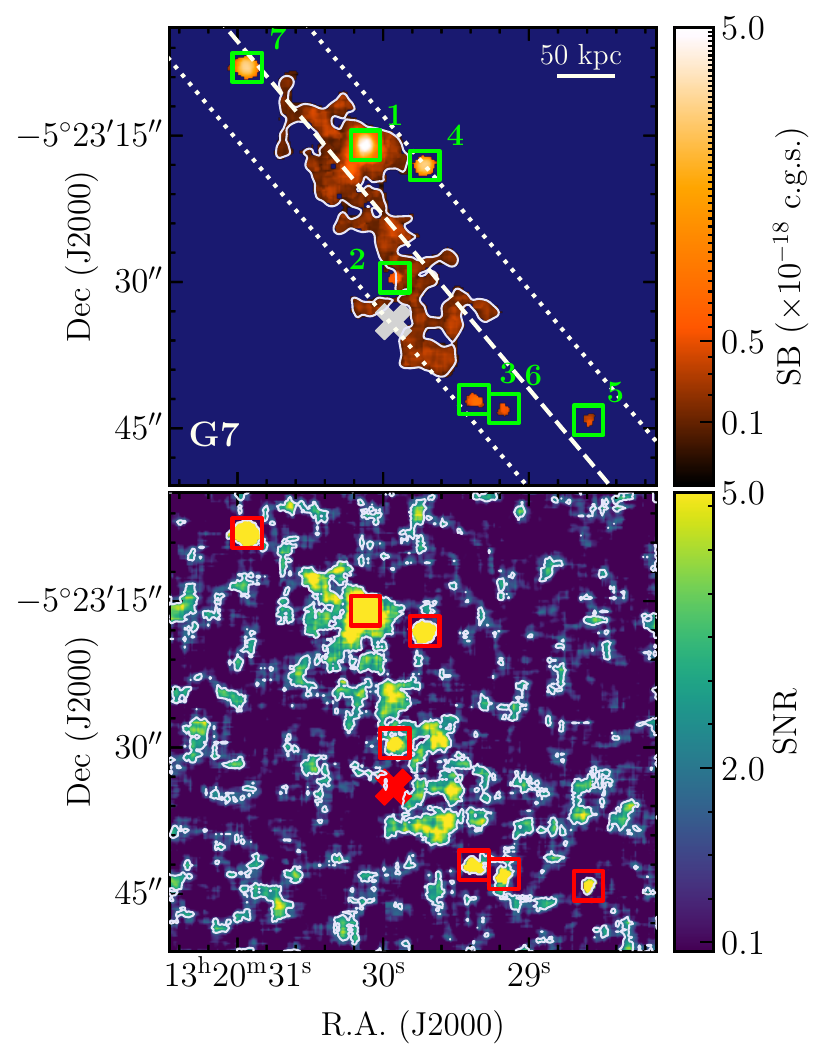}
    \caption{{\tt Top:} The surface brightness (SB) profile of the extended filament like structure as well as the associated LAEs of the G7 system. The contour indicates the SB $=10^{-19} \, \rm erg\, s^{-1}\, cm^{-2} \, arcsec^{-2}$. {\tt Bottom:} The SNR map of the extended emission and the associated LAEs, plotted on top of a single wavelength layer associated to the extended emission. The white contours correspond to the SNR of 2. In both panel, the positions of the seven LAEs are highlighted by the squares, while the quasar position is marked by a ``{\bf x}'' symbol. Maximum projected length of this structure is $\approx260$~pkpc. Refer to the text for the details about white dashed and dotted lines. 
    }
    \label{fig:filament}
\end{figure}

We reanalyzed the MUSE data using CubEx \citep{Cantalupo2019} on the quasar's point spread function (PSF) and continuum subtracted cube, focusing on 5535–5600~\ang\ ($\pm2000$~\kms\ from $z=3.577$). We searched for sources with $>3500$ connected voxels with SNR~$\geq 1.8$. To enhance sensitivity to low-SB sources, we applied a 4-pixel ($0.8''$) Gaussian spatial smoothing. This analysis revealed a large extended structure comprising $>10000$ connected voxels, with a projected linear size of $\approx260$~pkpc. We confirmed that the structure spans 16 distinct wavelength layers to ensure the detection is not spurious.

Fig.~\ref{fig:filament} displays the SB map (top) of the detected structure, with the 7 LAEs marked by green squares and (bottom) the SNR map of the same, overlaid on a single wavelength layer associated to the extended emission. The contours in the middle panel highlights the SB level of $10^{-19}\, \rm erg\, s^{-1}\, cm^{-2}\, arcsec^{-2}$, while that on bottom denotes $\rm SNR = 2$. Although faint, the detection is significant as it aligns closely with the LAE positions. Two LAEs (Ids: 1 and 2) lie directly within the contour, while two others (Ids: 4 and 7) are just outside it. The white dashed and dotted lines drawn on top of the SB-map indicate the best-fit linear alignment of the LAEs and their maximum deviation, $\epsilon$. The figure shows excellent correspondence between the extended emission and the filamentary structure traced by the LAEs.

The absence of emission at the quasar's location (marked by the {``{\bf x}''}) is likely due to enhanced noise from PSF subtraction. However, this background source allows direct measurement of the filament's \nh\ and $U$. Bayesian analysis shows $\log_{10}\,$\nh$/{\rm cm^{-3}}$ ranges between $-4.0$ and $-2.6$ \Rev{(see Section~\ref{sec:result_val}), corresponding to $\log_{10}\, U$ of $-0.8$ to $-2.2$ (for HM05)}. LAE Id: 2, located at 34~pkpc of the quasar sightline and $\approx 300$~\kms\ along the LOS, could enhance the local radiation field, raising the density estimate by $0.33 \pm 0.49$~dex \citep{Fumagalli2016}. Even with this correction, \nh\ remains low. 



\subsection{Origin(s) of the \lya\ nebula tracing the cosmic web}  

The projected linear size of approximately 260~pkpc classifies this extended emission as ``giant'' \lya\ nebula. Such giant \lya\ nebulae are generally detected around high-$z$ quasars \citep{Cantalupo2014, Borisova_2016} and between quasar pairs \citep{Tornotti2024, Herwig2024}. However this is the first detection of a giant \lya\ emission tracing cosmic filaments and associated with normal Lyman-$\alpha$ emitting galaxies  \Rev{\citep[see][for another recent example]{Tornotti2024filament}}. Note that none of these LAEs shows \HeII~$\lambda$1640 or any other emission line (such as \CIV). The low \lya\ emission equivalent width ($\approx50$~\ang\ in rest frame) and faint continuum also suggest the lack of AGN activity in these LAEs.

The primary source of this extended-\lya\ emission might be the ``in-situ'' recombination radiation following photoionization by the UV-photons. The \lya\ SB expected from a LLS illuminated by the HM05 UVB at $z\approx3.5$ is $2.2 \times 10^{-20}\, \rm erg\, s^{-1}\, cm^{-2} \, arcsec^{-2}$ \citep[see e.g.,][]{Cantalupo2005}, which is 5 times less than what we detect. Note, however, that the G7 system is $\approx 10$ times overdense compared to typical regions at this redshift. Consequently, UV photons from the seven LAEs likely contribute significantly to the emission. Their combined SFR, derived from UV continuum flux, is about four times higher than expected from the cosmic SFR density at this redshift \citep{Madau2014review}, within their comoving volume. This excess radiation field could explain the observed surface brightness. 


Without considering this excess radiation in overdense fields, \cite{Bacon2021} proposed that UV-faint galaxies could significantly contribute to the extended \lya\ emission. Their figure~14 suggests that a LF with a slope of $\lesssim -1.8$ integrated down to a \lya\ luminosity of 0 or $\approx 10^{37}\, \rm erg\, s^{-1}$, can result in a SB of $\approx 10^{-19}\, \rm erg\, s^{-1}\, cm^{-2} \, arcsec^{-2}$. Similarly, \cite{Guo2024} proposed that high-$z$ LAEs often have multiple satellite companions that might be the sources of the extended emission $\geq50$~pkpc. Although in our case a boost of the UV background by a factor similar to the galaxy overdensity is sufficient to explain the extended emission, we cannot exclude a contribution from fainter galaxies below the detection limit as this depends on the assumptions about the unknown faint end of the LAE luminosity function in such environments. Deeper observations and broader sky coverage with MUSE are essential for uncovering further insights into this intriguing cosmic structure.

\section*{ACKNOWLEDGEMENT}

\Rev{We thank the anonymous referee for their useful suggestions.} We thank Marijke Segers, Lorrie Straka, and Monica Turner for their early contributions to the MUSEQuBES project. EB and SM thank Raghunathan Srianand for useful suggestions. This work has used IUCAA HPC facilities. We gratefully acknowledge the European Research Council (ERC) for funding this project through the Indo-Italian grant. We thank Vikram Khaire and Abhisek Mohapatra for useful discussions. This paper uses the following software: NumPy \cite[]{Harris_2020}, SciPy \cite[]{Virtanen_2020}, Matplotlib \cite[]{hunter_2007} and AstroPy \cite[]{Astropy_2013}.

\bibliography{zbib}{}
\bibliographystyle{aasjournal}

\end{document}